\documentstyle[prl,aps,multicol]{revtex}
\input{epsf}
\begin{document}
\draft
\title{Quantum Poincar\'e Recurrences }                      

\author{Giulio Casati$^{(a,b,c)}$, Giulio Maspero$^{(a,b,c)}$ and 
Dima L. Shepelyansky$^{(d,*)}$}
\address{$^{(a)}$Universit\`a di Milano, sede di Como, Via Lucini 3,
22100 Como, Italy}
\address{$^{(b)}$Istituto Nazionale di Fisica della Materia, 
Unit\`a di Milano, Via Celoria 16, 20133 Milano, Italy}
\address{$^{(c)}$Istituto Nazionale di Fisica Nucleare, Sezione di Milano,
Via Celoria 16, 20133 Milano, Italy} 
\address {$^{(d)}$  Laboratoire de Physique Quantique, UMR 5626 du CNRS,
Universit\'e Paul Sabatier, F-31062 Toulouse Cedex 4, France}

\date{\today}
\maketitle
\begin{abstract}
We show that quantum effects modify the decay rate of Poincar\'e 
recurrences $P(t)$ in classical chaotic systems with hierarchical 
structure of phase space. 
The exponent $p$ of 
the algebraic decay $P(t) \propto 1/t^p$ is shown to have the 
universal value $p=1$ due to tunneling and localization effects.
Experimental evidence of such decay should be observable in mesoscopic
systems and cold atoms.
\end{abstract}

\begin{multicols}{2}
\narrowtext

The general structure of classical phase space in chaotic Hamiltonian 
systems displays a hierarchical mixture of integrable and chaotic 
components down to smaller and smaller scales \cite{Licht}.
This complicated structure leads, in particular,
to an anomalous power law decay of Poincar\'e recurrences $P(t)$ and
correlations $C(t)$ inside the chaotic components \cite{cs1,kern}.
Physically, such slow decay appears due to a decrease, down to
zero, of the diffusion rate
for a trajectory when it approaches the chaos border determined
by some critical invariant curve \cite{cs1,kern,Ott,chrn,ruffo}.
Typically, $P(t) \sim C(t)/t \sim t^{-p}$ with $p \approx 1.5$ \cite{cs1}.
As a result the integrated correlation function, which determines the
diffusion rate ($D \sim \int C dt$), can diverge thus leading to a 
super-diffusive propagation \cite{chizetp}. 
Such effects are important for electron
dynamics in super-lattices
where usually the phase space has a mixed structure \cite{geisel}.

The above anomalous properties had been studied in great detail for classical
systems \cite{cs1,kern,Ott,chrn,ruffo,chizetp,geisel}. However, the question
how they are affected by quantum dynamics was not addressed up to now.
This problem becomes more and more important not only 
due to its fundamental nature but
also in the light of recent experiments with mesoscopic systems. Indeed,
different types of ballistic 
quantum dots can now be studied in laboratory experiments \cite{marcus} 
and  the phase space 
in such systems generally has a mixed structure. Since, the probability to
stay in a given region is directly related with $P(t)$ and $C(t)$,
its  slow decay can significantly affect conductance properties. In
particular it has been proposed that such decay should lead to fractal 
conductance fluctuations \cite{ketz}, the experimental observation of which 
has been reported recently \cite{ketz2}. According to \cite{ketz,ketz2} 
the fractal exponent $\sigma$ for conductance fluctuations is directly related
to the exponent $p$ as 
$\sigma = 2-p/2$.

A different type of  systems in which such effects should be
observable experimentally is given by cold atoms in external laser fields
where the Kicked Rotator model of quantum chaos has been  built experimentally
\cite{reiz,am}.
Possibilities of experimental investigation of slow probability decay in such
systems has been discussed recently \cite{reiz2}.

The experimental studies of slow power correlation decay 
in the regime of quantum chaos
are also important from the fundamental point of view, since here the 
typical scale of correlation decay is much larger than the Ehrenfest
time scale $t_E \sim \ln 1/\hbar$ on which the minimal coherent
wave packet spreads over the avaible phase space. To the best of our 
knowledge the comparison of classical and quantum correlations in such
regime has not been investigated so far. Only recently such comparison has
been made in the regime of hard chaos with exponential correlation decay 
\cite{second}.
In this paper we address directly the comparison between classical and quantum
Poincar\'e recurrences (QPR), related to the correlations decay,
in the regime with mixed phase
space. Our results demonstrate a new universal law for QPR related to
localization and tunneling effects.

To investigate the QPR we use the model of kicked rotator with
absorbing boundary conditions studied in \cite{Borg,second}.
The evolution operator over the period $T$ of the perturbation is given by
\begin{eqnarray} 
\label{qmap}
\bar{\psi} = \hat{U} \psi = \hat{{\cal{P}}} e^{-iT\hat{n}^2/4} e^{-ik
\cos{\hat{\theta}}}
 e^{-iT\hat{n}^2/4} \psi,
\end{eqnarray}
where $\hat{{\cal{P}}}$ is a projection operator 
over quantum states $n$ in the
interval $(- N/2, N/2)$.
Here, we put $\hbar=1$ so that
 the commutator is $[\hat{n},\hat{\theta}]=-i$ and the
classical limit corresponds to $k \rightarrow \infty$, 
$T \rightarrow 0$ while
the classical chaos parameter $K=kT$ remains constant.
In the classical limit the dynamics is described by the Chirikov standard map:
\begin{eqnarray} 
\label{cmap}
\bar{n} = n + k \sin{ [ \theta + { T n \over 2} ]}; \ 
\bar{\theta} = \theta + {T \over 2} (n+\bar{n}),
\end{eqnarray}
in which orbits are absorbed outside the interval $-N/2 < n < N/2$.
In order to study the classical and quantum survival probability
$P(t)$ we fixed the ratio $N/k=4$ and take 
the classical chaos parameter $K=2.5$,  so 
that the classical phase space has a hierarchical structure of 
integrable islands and chaotic components.

A typical example 
of classical and quantum survival probability decay is shown in
Fig.1. The classical probability $P(t)$ decays with a power law with
exponent $p \approx 2$ in a range of six orders of magnitude.
In this case the exponent is slightly different from the typical value
$1.5$. Indeed as discussed in \cite{cs1,kern,Ott,chrn} the
exponent can vary from system to system (and even oscillates
with $\ln t$) depending on the
local structure of phase space in the vicinity of the critical 
boundary invariant curve whose rotation number can play an important role.

The quantum survival probability $P_q(t)$ is plotted for different
values of $N$ ( effective $\hbar$ is proportional to $1/N$) in Fig.1. 
In agreement with the correspondence 
principle, the quantum probability follows the classical value during a rather 
long time scale which grows with $N$. For longer times the quantum 
probability $P_q(t)$ is found to approximately follow the quantum decay law
\begin{equation}
\label{slp1}
P_q(t) \approx C / t
\end{equation}
where $C$ is some constant.
This latter decay continues during a rather long interval 
of time (4 orders of magnitudes in time
for the case $N=3^6$). Of course (\ref{slp1}) 
represents an intermediate 
asymptotic behavior since for $t \rightarrow \infty$ the decay will be exponential $P_q(t) \sim \exp{(-\Gamma_{min} t)}$ where $\Gamma_{min}$ is
determined by the minimal imaginary part of the eigenvalue $\lambda$ of the
evolution operator $\hat{U}$  $(\hat{U} \psi_{\lambda} = e^{-i\lambda} 
\psi_\lambda)$.

In order
to better understand the origin of quantum 
behavior (\ref{slp1}) we studied the probability decay in a
simpler case where the classical dynamics is completely chaotic ($K=7$) or
quasi-integrable ($K=0.5$).
The quantum and classical probability decays are shown in Fig.2
for  $k=5$ and absorption for $n \leq 0 $ and $n > N=500$.
Initially the probability is concentrated at $n=0$.
For $K=7$ the classical probability decays, asymptotically, 
exponentially with time $P(t) = 0.11 \exp{(-\gamma t)}$.
The value of $\gamma$ can be found from the solution of the Fokker-Planck
equation with absorbing boundary conditions which gives
$\gamma \approx {D \pi^2 / 2 n^2}$, where $D = \beta k^2/2$ is the
classical diffusion rate (see eq.(4) in \cite{Borg}). The value of
$\beta$ depends on the classical chaos parameter and for $K=7$ is
$\beta \approx 2.8$ \cite{Borg}.
Therefore the expected theoretical value is $\gamma = 6.9 \cdot 10^{-4}$
which is close to the numerical value $\gamma = 6.4 \cdot 10^{-4}$.
The asymptotic exponential decay starts after the diffusive time
$t_D \approx 1/\gamma$.
The quantum probability in this case (Fig.2) follows the 
classical one during some interval of time after
 which it decays according to eq.(\ref{slp1}). 
 
The quantum decay (\ref{slp1}) can be understood in the following way.
The quantum eigenstates are localized with localization
length $\ell \approx D/2$ \cite{loc}.
Therefore  the absorption time for a state at a distance $n$ from the 
absorbing boundary is $t \sim \exp{(2n/\ell)}$. Since $n$ is
proportional to the total measure $\mu \sim n \propto \ell \ln t$,
it follows that the survival probability is $P_q(t)
 = d \mu / dt \sim \ell / t$ in
agreement with (\ref{slp1}).
The same estimate can be obtained by expanding an initial state over
eigenstates of the evolution operator (\ref{qmap}) with probabilities 
$ \left| c_n \right|^2 \sim {e ^{- 2n/\ell} / \ell}$
 and ionization rates 
$\Gamma_n \sim  \left| c_n \right|^2 \sim {e ^{- 2n/\ell} /D}$.
Then the survival probability 
$P_q(t) \sim \int_0^{\infty} \left| c_n \right|^2 e^{-\Gamma_n t} dn \sim
{D/t}$.
Our data at $K=7$ for different values of $k$ $(k=5, 6, ... 10)$
give that $C = a D^b$, with $a= 0.27(5)$ and $b=0.92(13)$.

The above derivation of the decay law (\ref{slp1}) refers to
the regime of quantum localization of chaos.
For the integrable case one still has the rates 
$\Gamma_n \sim  \left| c_n \right|^2 \sim {e ^{- n/\ell_{eff}} }$
where $\ell_{eff}$ is an effective length determined by tunneling in the
classically forbidden region. However here the fluctuations in time are
stronger since $\ell_{eff}$ depends on the  local structure of invariant
curves and islands in the integrable domain. Nevertheless the global
decay at $K=0.5$ is in agreement with the $1/t$ law (see Fig.2). 
In both cases $(K=7; 0.5)$, the quantum
decay proceeds in a much slower way than  the corresponding classical one.
Due to the above reasons even in the case of mixed phase space the decay
follows the $1/t$ law in accordance with numerical data of Fig.1.
The $1/t$ behavior can continue up to a time $t_{max} \propto 
\exp{(N/\ell_{eff})} \sim 1/\Gamma_{min}$ 
which is determined by the minimal decay rate in the system.
For $t > t_{max}$ the quantum survival probability $P_q(t)$ decays 
exponentially with time $ P_q(t) \propto \exp{( -t/t_{max})}$.

The ionization rate $\Gamma_{\lambda}$ of an eigenstate  $\psi_{\lambda}$
localized at a distance $n$ from the absorbing boundary is proportional to 
$\Gamma_\lambda \propto \exp{(-n/\ell_{eff})}$, therefore the
number of such states is proportional to the measure 
$\mu \sim n \sim \ln 1/\Gamma$. As a result the probability to find a value
$\Gamma$ in the interval $d\Gamma$ is $dW/d\Gamma \sim d\mu/d\Gamma \sim
1/\Gamma$. Our numerical data for $dW/d\Gamma$ obtained 
in the localized regime confirm this $1/\Gamma$ dependence (see \cite{tobe}).

An interesting question is at what time scale the QPR start to deviate
from their classical behavior. To test these deviations we determined
them in two different ways. The first one is defined as the time $t_q$
at which the ratio of classical over quantum probability is 
$P(t_q) / P_q(t_q) = 0.9$. This condition gives the first
quantum deviation and was studied in \cite{second} where
it was found $t_q \propto \sqrt{N} \propto \sqrt{1/\hbar}$. This time scale
can be explained on the basis of analysis of 
eigenvalue fluctuations 
of the evolution operator $\hat{U}$ in the complex plane or as
the result of weak localization corrections \cite{second}.
Our numerical data shown in Fig.3 confirm this dependence even if the
situation is qualitatively different from that in \cite{second},
where the integrable component was absent, while here
we have power law decay in a mixed phase space (see Fig.1).

However at time $t_q$ quantum probability only starts to deviate from
classical one while the behavior (\ref{slp1}) 
can set in after a longer time $t_H$. We determine this time in two ways:
i) $t_H \approx \ln P(t_q) / \ln P_q(t_q) = 0.9$ (circles and stars in Fig.3);
ii) as the crossing point between the two extrapolations for the
classical behavior
$P(t) \propto 1/t^2$ and the quantum one $P_q(t) \propto 1/t$
(crosses in Fig.3). The data for stars in Fig.3 
were obtained by averaging over $50$ values of $N$ near a given value.
We used two methods to determine the time scale $t_H$ because the first
definition gives large oscillations related to the oscillations of the 
classical $P(t)$ in $\ln t$ while the second definition allows to 
smooth out the effect of the oscillations.
The numerical data are in agreement with the dependence $t_H \sim N \propto
1/\hbar$. This time can be interpreted as the Heisenberg time scale,
which is determined by inverse level spacings, and after which the quantum 
behavior becomes qualitatively different from the classical one.

The above scales $t_q$ and $t_H$ can be also seen in the quantum evolution 
of the Husimi function 
obtained from the Wigner function 
by smoothing over the size of $\hbar$ cell. This evolution is
presented in Fig.4. For short times $t < t_q = 150$ the classical
and the quantum probability distributions in the phase space $(n,\theta)$
look rather similar. The quantum Husimi function reflects the underline
fractal structure of the classical distribution. For larger times the
classical recurrencies are determined by more and more fine scales in the 
phase space and the classical distribution becomes localized
around small islands near critical invariant curves. On such long
times $t > t_H =3000$ the quantum evolution in phase space 
in proximity of critical invariant curves becomes influenced
by quantum interference effects that results in a qualitatively different
probability distribution (Fig.4). The penetration of quantum probability
in phase space on smaller and smaller scales proceeds via a very slow tunneling
process and as a result the Husimi function remains almost 
unchanged when the time is increased almost by two order 
of magnitudes (Fig.4).

At very long times $t > t_{max}$ the quantum decay becomes exponential; this
corresponds to eigenstates which are localized in the center of the island 
(Fig.4). We have been able to observe 
such localized asymptotic Husimi function for $N=3^4$ (Fig.4).
However, already for $N=3^5$ the time $t_{max}$ is too large to reach it in our
numerical simulations where we followed 
the evolution up to $3 \cdot 10^7$.
Apparently this time is already comparable with $t_{max}$ and this
explains the
strong oscillations of $P_q(t)$ for $t > 10^6$ (Fig.1). At this time
the quantum probability penetrates inside four small islands while at
$t \approx 1.5 \cdot 10^7$ the probability becomes
concentrated inside the main central island.
We would like to stress that $t_{max}$ grows exponentially with $N$
($N \sim 1/\hbar$). As the result,
the quantum decay $1/t$ in semi-classical region proceeds up to
enormously long times.
In this regime the correlation functions practically do not
decay because $C(t) \propto t P(t) \approx const$.

In conclusion, we established a new universal law for probability and
correlations decay in quantum systems both in 
the exponentially localized regime and in 
regimes in which classical dynamics has a mixed phase space.
It should be possible to observe this universal behavior in the 
kicked rotator model which has been studied in experiments 
with cold atoms \cite{reiz,am}.
Similar effect should be also observable in the experiments with micrometer
droplet lasing discussed recently in \cite{Stone}.
The quantum decay (\ref{slp1}) of $P_q(t)$  has exponent $p=1$ that should also 
effect the fractal dimension
of conductance fluctuations in soft billiards studied in \cite{ketz,ketz2}.
According to these results the fractal dimension of these fluctuations
should be $\sigma = 2 - p/2 = 1.5$.
This regime should start at large times $t > t_H$ which requires
to analyze conductance fluctuations with very fine resolution of magnetic
field. The above value of $\sigma$ has not been seen in experiments 
\cite{ketz} that apparently indicates that the experimental resolution was
not yet sufficient to detect quantum Poincar\'e recurrences
at large times.

\begin{figure}
\centerline{\epsfxsize=9cm \epsfbox{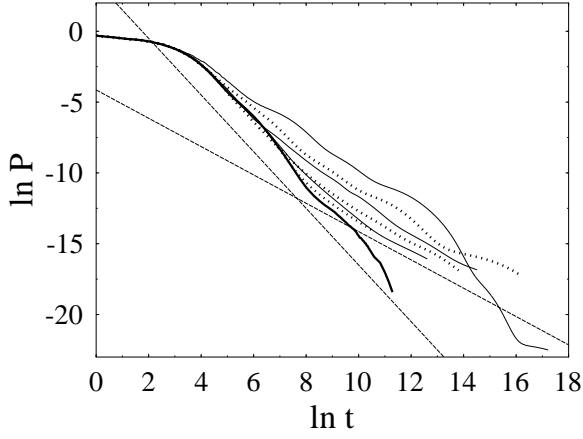}}
\vspace{4mm}                                 
\caption{
Classical (thick continuous line) and quantum 
(thin continuous and dotted lines)
probability decays for $K=2.5$ and $N/k=4$. The two  dashed  straight lines
show slope $2$ and $1$. The quantum curves correspond to $N=3^p$ with $p$ 
increasing from $5$ (upper continuous) to $10$ (lower dotted).
The starting conditions are two symmetric lines at $n=\pm N/3$, both for the  
classical and quantum evolution.
}
\end{figure}

\begin{figure}
\centerline{\epsfxsize=9cm \epsfbox{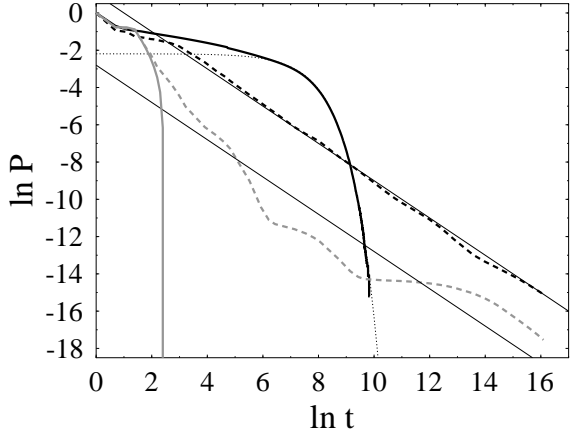}}
\vspace{4mm}
\caption{ 
Classical (continuous lines) and quantum (dashed lines)
probability decays for $K=0.5$ (grey lines) and $K=7$ (black lines); 
parameter $k=5$.
The evolution starts 
 at $n=0$ and  the probability is absorbed for $n \leq 0$ and $n > 500$.
The two thin straight  lines have slope one, while the dotted grey curve  
shows the fit for the classical decay at $K=7$:
 $P = 0.11 \exp{( - \gamma t)}$,
$\gamma= 6.4 \cdot 10^{-4}$.
}
\end{figure}

\begin{figure}
\centerline{\epsfxsize=9cm \epsfbox{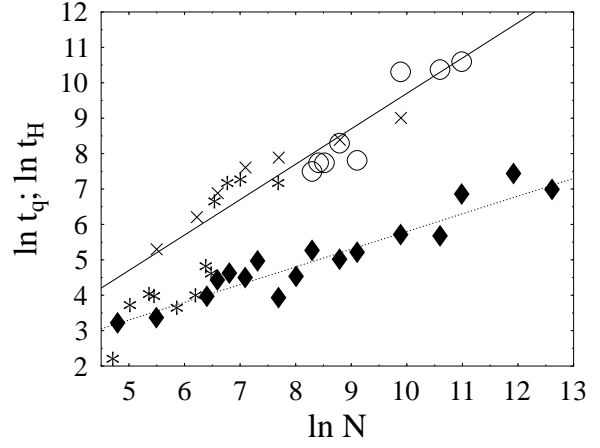}}
\vspace{4mm}                          
\caption{
Dependence of quantum time scales $t_q$ (full diamonds) and $t_H$
(circles, crosses, stars) on the system size $N$ (see the text). The
dotted and continuous straight lines show the theoretical slopes $1/2$ and $1$,
for $t_q$ and $t_H$ respectively.
}
\end{figure}

\begin{figure}
\centerline{\epsfxsize=9cm \epsfbox{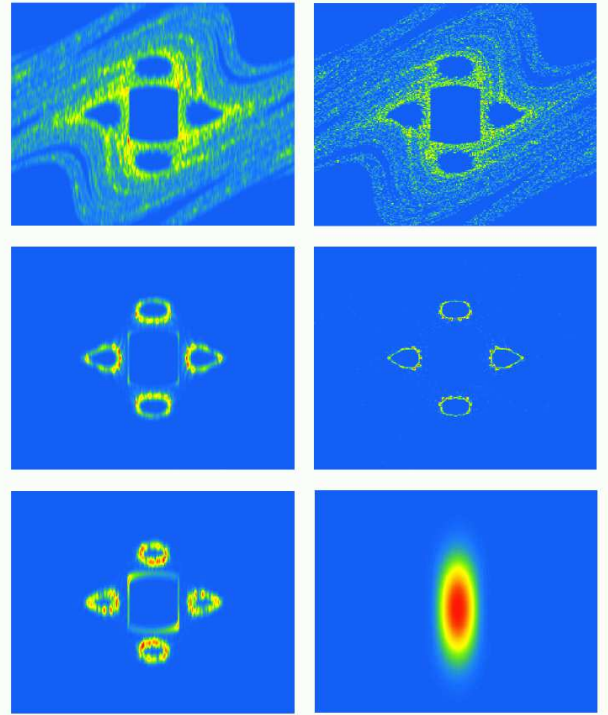}}
\vspace{4mm}
\caption{       
Left column: Husimi function  at time
$10^2$ (top), $5 \cdot 10^3$ (middle) and $3 \cdot 10^5$ (bottom), for 
the case in Fig.1 with $N=3^8$.
Right column: classical density plot  
of orbits surviving up to time $10^2$ (top) and  $5 \cdot 10^3$ (middle).
The right bottom shows Husimi function for $N=3^4, t=10^7$.
Distributions are averaged in a small time interval $\delta t = 20$ near
the given $t$ values.
The size of the phase region ($\theta$,$n$) is:
 $0 \leq \theta \leq 2 \pi$, $-N/2 \leq n 
\leq N/2$. The color is proportional to the density:
blue for zero and bright red for maximal density
(the color scale is the same for classical and quantum cases).
}
\end{figure}

\end{multicols}

\end{document}